# РАДУГА ДЕКАРТА-НЬЮТОНА-ЮНГА
Алексей Панов

Радуга – это гигантский оптический эксперимент, который демонстрирует нам природа всякий раз, как только складываются подходящие условия: уходящая от нас стена дождя освещается низко стоящим солнцем, и в небе вспыхивают огненные дуги. Основной вклад в физическую теорию радуги был сделан Декартом.

## Декартова радуга

В 1637 году появился знаменитый трактат Декарта "Рассуждение о методе, чтобы хорошо направлять свой разум и отыскивать истину в науках". В многочисленных приложениях к трактату Декарт продемонстрировал мощь своего метода, и одна из этих демонстраций – это описание механизма образования радуги.

**Эксперимент**. С помощью своего метода Декарт быстро устанавливает, что для понимания радуги сначала нужно изучить взаимодействие пучка солнечных лучей с отдельной каплей. В качестве такой гигантской капли он берет круглый стеклянный сосуд, наполненный водой. Декарт наблюдает за сосудом, освещенным солнцем, он обходит вокруг сосуда, обносит его вокруг своей головы, и выясняет, что в направлении к солнцу эта освещенная гигантская "капля" излучает целый конус световых лучей с углом полураствора 42° (рис. 1). Уже этого одного факта достаточно, чтобы понять механизм образования радуги.

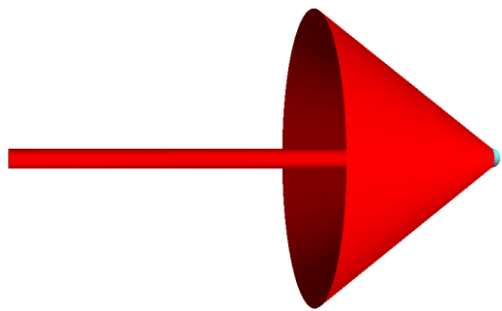

Рис. 1. От солнца на каплю приходит пучок параллельных лучей, в ответ капля излучает целый конус лучей с углом полураствора 42°

**Механизм образования радуги**. Итак, уходящая стена дождя освещается появившимся солнцем. Давайте начнем с самой простой картинки, посмотрим, что происходит в вертикальной плоскости, проходящей через солнце и глаз наблюдателя (рис. 2).

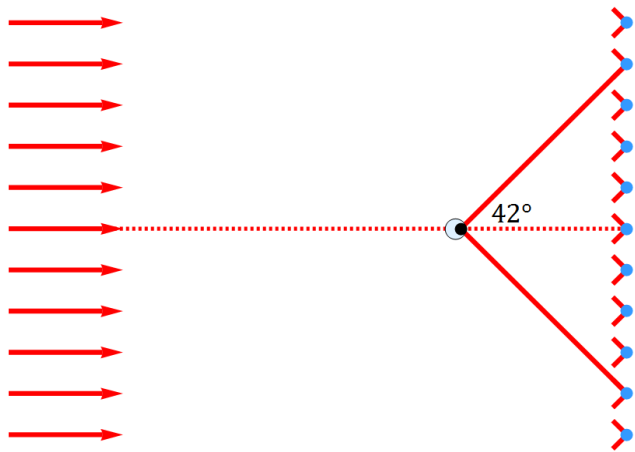

Рис. 2. Стена дождя, лежащая в плоскости, перпендикулярной плоскости рисунка, освещается пучком параллельных лучей, приходящим от солнца

Согласно Декарту, каждая капля излучает в направлении к солнцу целый конус световых лучей. Для капель, лежащих в плоскости рисунка, этот конус представлен двумя лучами: один идет вверх от капли, другой – вниз, оба под углом 42° (рис. 2). Из всех этих лучей в глаз наблюдателя попадают лучи только от двух капель (рис. 2).

Если же говорить обо всем пространстве в целом, то в глаз наблюдателя попадают лучи от тех капель, которые он сам видит под углом 42° относительно оси солнце-глаз. Эти лучи лежат на конусе с углом полураствора 42°. Наблюдатель воспринимает их, как яркую окружность с угловым радиусом 42°, удаленную от него на некоторое расстояние. Это и есть радуга, пока еще не цветная и не обладающая шириной, но, тем не менее, радуга. Рисунок 2 – самый важный в теории радуги и с ним нужно хорошенько разобраться.

**Упражнение 1**.
1. Убедитесь, что вид радуги (ее угловой размер) не зависит от расстояния между наблюдателем и стеной дождя.
2. Убедитесь, что капли, не лежащие в одной плоскости, а заполняющие некоторую область в пространстве, все равно будут создавать видимую наблюдателю радугу тех же размеров.

На рисунке 2 не хватает еще одной существенной детали – поверхности земли, ограничивающей обзор реального наблюдателя. Посмотрите на рис. 3, где добавлена эта поверхность и ответьте еще на несколько вопросов.

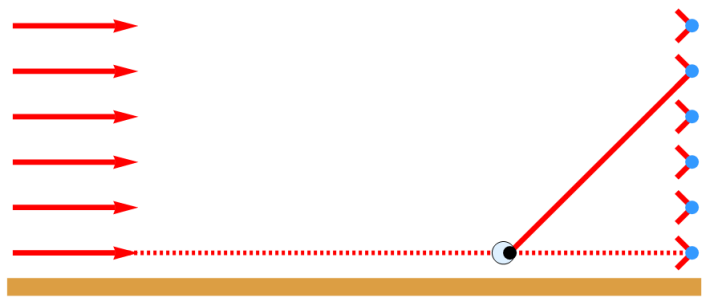

Рис. 3. Поверхность земли ограничивает обзор наблюдателя

**Упражнение 2**.
1. Какое время суток представлено на рис. 3? Какую часть радуги, видит наблюдатель, изображенный на рис. 3?
2. Как будет меняться вид радуги, если наблюдатель будет постепенно подниматься вверх на аэростате?
3. Как будет выглядеть радуга, если изменится высота солнца над горизонтом? Будет ли видна радуга, если солнце поднимется достаточно высоко?

**Лучи, отвечающие за образование радуги.** Солнечные лучи могут по-разному взаимодействовать с каплей. Одни отражаются от передней поверхности капли. Другие, преломляясь, входят в каплю, несколько раз отражаются внутри нее и только потом выходят наружу.

С помощью дополнительных экспериментов, а именно с помощью экранирования световых лучей, Декарт установил, что за образование радуги отвечают те лучи, которые входят в каплю, отражаются от ее задней поверхности и затем выходят из нее (рис. 4).

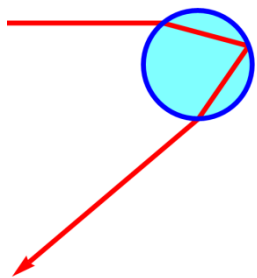

Рис. 4. Луч однократно отражается внутри капли, именно такие лучи отвечают за образование радуги

Но дальше Декарт пишет, что он так и не понял, почему выходящие из капли лучи образуют конус с углом в 42°. Не понял, "*пока не взялся за перо и не вычислил ход всех лучей, которые падают на различные точки водяной капли, чтобы узнать под каким углом они могут попасть в глаз после двух преломлений и одного отражения*".

Чтобы произвести расчет траектории таких световых лучей, нужно знать закон преломления, ведь световой луч преломляется два раза, один раз на входе в каплю и другой на выходе (рис. 4).

**Закон преломления**. Декарт был одним из тех, кто открыл закон преломления. Во всяком случае, он опубликовал его в тех же самых "Рассуждениях о методе". И вне всяких сомнений, он был первым, кто применил его для объяснения радуги.

Закон преломления, сформулированный Декартом, можно записать в виде
$$n_1 \sin i = n_2 \sin r.$$
Здесь $n_1$ – это показатель преломления среды, из которой приходит световой луч, а $n_2$ – показатель преломления среды, в которую он преломляется.

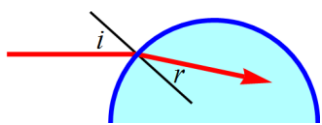

Рис. 5. Преломление светового луча при переходе из одной оптической среду в другую

Угол $i$ – это угол между направлением приходящего светового луча и перпендикуляром к поверхности преломления, он называется *углом падения*, а $r$ – это угол между преломленным лучом и тем же самым перпендикуляром, он называется *углом преломления* (рис. 5). При этом падающий луч, перпендикуляр и преломленный луч лежат в одной плоскости.

**Преломление на капле**. В нашем случае капля расположена в воздухе, показатель преломления которого с большой точностью можно считать равным единице, $n_\text{воздуха} = 1$ и пока примем, что $n_\text{воды} = 4/3$. Перпендикулярами к поверхности капли служат ее радиусы, а все необходимые углы обозначены на рисунке 6.

Дополнительно, угол между выходящим из капли лучом и первоначальным направлением падающего на каплю светового луча там обозначен $\theta$, будем называть его *углом выхода*.

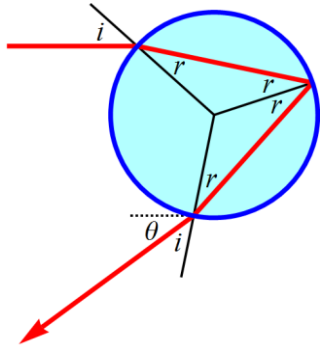

Рис. 6. Два преломления и одно отражение, $\theta$ – угол выхода

В этих обозначениях закон преломления на входе в каплю записывается в виде
$$\sin i = \frac{4}{3}\sin r,$$
на выходе имеем
$$\frac{4}{3}\sin r = \sin i,$$
что по сути одно и то же. И еще, в точке отражения на рисунке 6 учтено, что угол падения равен углу отражения.

Вот несколько фактов, полезных при расчете траектории светового луча.

**Упражнение 3**.
1. Убедитесь, что из закона преломления для капли следует, что
$$r = \arcsin\left(\frac{3}{4}\sin i\right).$$
2. Используя рисунок 6, докажите следующую формулу для угла выхода $\theta$
$$\theta = 4r - 2i.$$

**"Взялся за перо и вычислил".** Используя закон преломления, Декарт рассчитал ход 27 световых лучей, и этого ему оказалось достаточно, чтобы понять, почему капля излучает в направлении к солнцу конус с углом полураствора 42°. Посмотрим, что именно вычислял Декарт и как он интерпретировал свои вычисления.

Будем считать, что капля единичного радиуса освещается пучком параллельных солнечных лучей. Среди них есть центральный луч, который направлен в центр капли. Для любого луча расстояние от него до центрального обозначим $h$ и назовем эту величину *высотой входа* в каплю.

**Упражнение 4**. Проверьте, что для капли единичного радиуса высота входа луча $h$ и угол его падения $i$ связаны соотношением $h = \sin i$.

Для заданной высоты входа $h$ Декарт вычислял угол выхода $\theta$. Сначала Декарт провел вычисления для высот от 0,1 до 1 с шагом 0,1. После этого для высот от 0,81 до 0,98 он проделал те же вычисления с уменьшенным шагом 0,01. Таким образом, Декарт установил, что максимальный угол выхода $\theta$ достигается для луча с высотой $h = 0{,}86$ и этот угол как раз составляет $42°$.

Посмотрим, как Декарт геометрически интерпретировал этот результат. Компьютер позволяет не экономить на вычислениях, так что запустим сразу 101 луч и поглядим, что с ними происходит при выходе из капли. Сделаем это в два приема. Сначала на каплю единичного радиуса запустим 87 параллельных лучей, при этом начальный проходит через центр капли, последний входит в нее на высоте $0{,}86$. На рисунке 7 входящие лучи слились в короткую горизонтальную полоску. Внутри капли каждый луч один раз отражается, как на рисунке 4. Это здесь не нарисовано, а вот все выходящие лучи изображены. Мы видим, что когда высота входа луча растет от $0$ до $0{,}86$, угол выхода луча $\theta$ возрастает от $0°$ до $42°$.

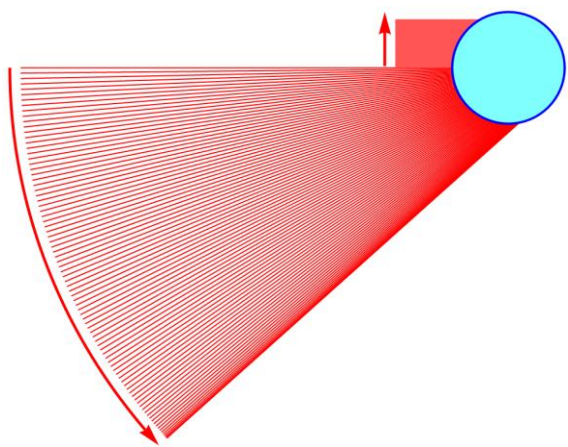

Рис. 7. По мере роста высоты входа лучей от $0$ до $0{,}86$
угол выхода увеличивается от $0°$ до $42°$

При этом сначала угол выхода растет равномерно, потом угол между соседними лучами начинается уменьшаться и при приближении к $42°$ становится нулевым – выходящий луч останавливается. Останавливается, и начинает двигаться в обратном направлении, что видно на рисунке 8, где высота входа лучей увеличивается от $0{,}86$ до $1$ и угол выхода уменьшается в три раза, от $42°$ до $14°$.

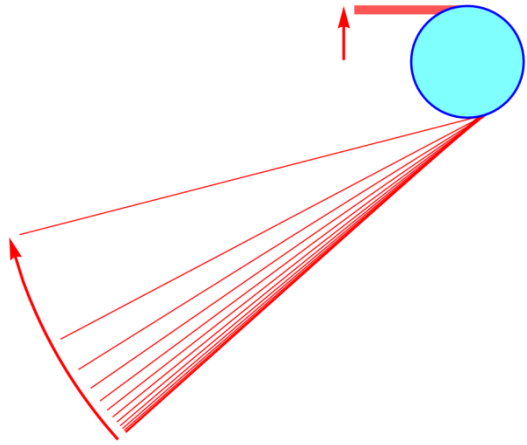

Рис. 8. Высота входа лучей растет от 0,86 до 1,
угол выхода уменьшается от 42° до 14°

В любом случае, вблизи крайнего луча, идущего под максимальным углом 42°, наблюдается высокая концентрация выходящих из капли световых лучей.

Уберем входящие лучи на рисунках 7 и 8, после этого наложим рисунки друг на друга. Получим все лучи, выходящие из нижней половины капли, а затем дополним их еще лучами, выходящими из верхней половины (рис. 9).

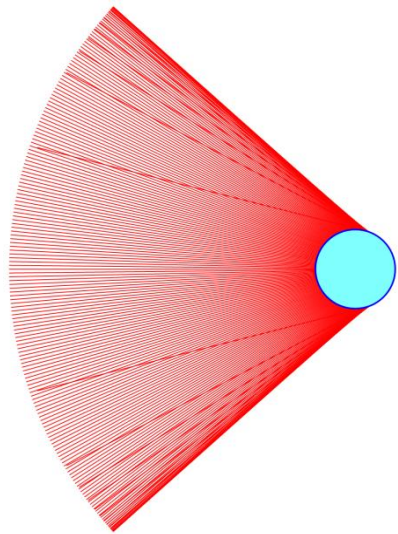

Рис. 9. Все лучи, выходящие из капли после одного внутреннего отражения

И теперь можно сказать, что концентрация лучей, выходящих из капли, мала вблизи центрального луча и чрезвычайно высока на границах пучка. Эти границы почти прямолинейны и практически совпадают с крайними лучами, наклоненными под максимальным углом в 42°.

Чтобы получить пространственное распределение лучей, необходимо провернуть весь рисунок 8 вокруг центрального луча. И теперь мы видим, что экспериментальное представление о капле, излучающей конус в направлении солнца, зафиксированное на рисунке 1, нуждается в уточнении. На самом деле, внутренность конуса на рисунке 1 целиком заполнена выходящими из капли лучами. На поверхности конуса с полураствором $42°$ наблюдается лишь *повышенная концентрация* световых лучей. Это существенное уточнение реальной картины взаимодействия солнечных лучей с каплей дождя. Тем не менее, глаз наблюдателя регистрирует именно эту повышенную концентрацию световых лучей. Поэтому предложенный ранее механизм образования радуги безусловно остается в силе.

Вычисления Декарта, которые мы обсудили в этом разделе, и его интерпретация этих вычислений составляют основу теории радуги.

**Упражнение 5**. Используя упражнения 3 и 4, докажите что для капли единичного радиуса
$$\theta(h) = 4\arcsin\left(\tfrac{3}{4}h\right) - 2\arcsin h.$$
Постройте график этой функции. Докажите, что ее максимум достигается при $h = 2\sqrt{15}/9$, а максимальное значение этой функции равно $42°2'$.

**Как она выглядит на самом деле?** Пора взглянуть на настоящую радугу. В отличие от радуги Декарта она широкая и цветная (рис. 10), но это мы обсудим чуть позже. А пока посмотрим на тени от деревьев. На земле они параллельны и так же, как параллельные рельсы, сходятся в одной точке, благодаря перспективе. Эта точка расположена на горизонте под самой вершиной радуги. А если для каждого дерева провести луч от его вершины к вершине его тени, то все эти лучи сойдутся в самом центре круга, который ограничивает радуга.

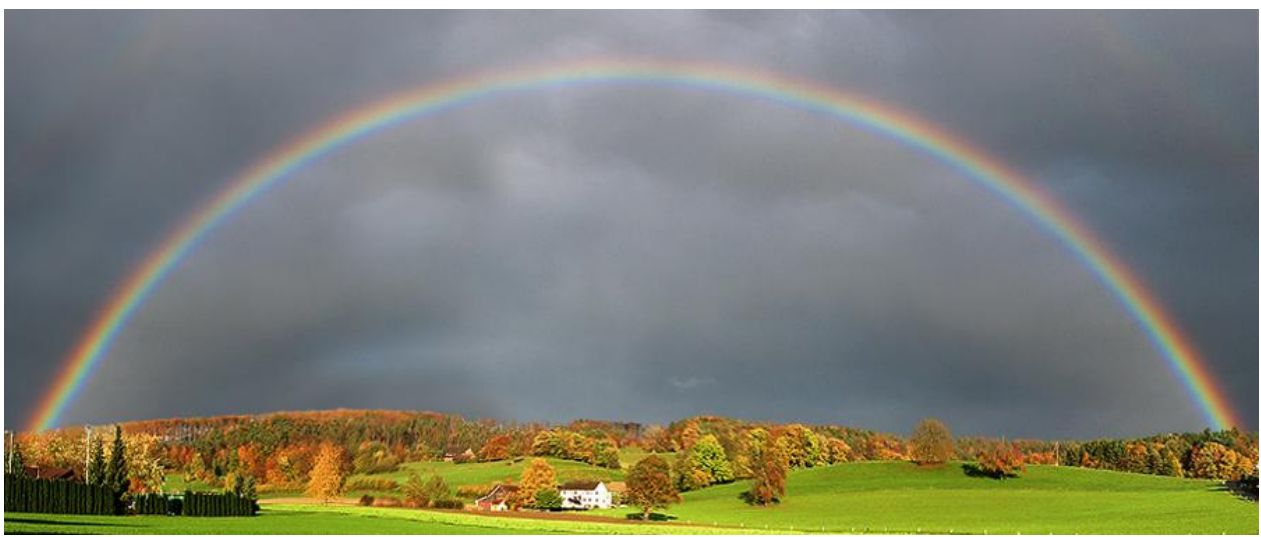

Рис. 10. Радуга

А теперь представьте себе, что вы находитесь внутри этой картины. Посмотрите сначала на тень своей головы, а потом на любую точку радуги. Угол между этими двумя направлениями как раз составит 42°, вычисленные Декартом.

А теперь еще посмотрите на верхние уголки рисунка 10 – может, вы разглядите там неяркие цветные полосы. Если нет, взгляните на рисунок 11.

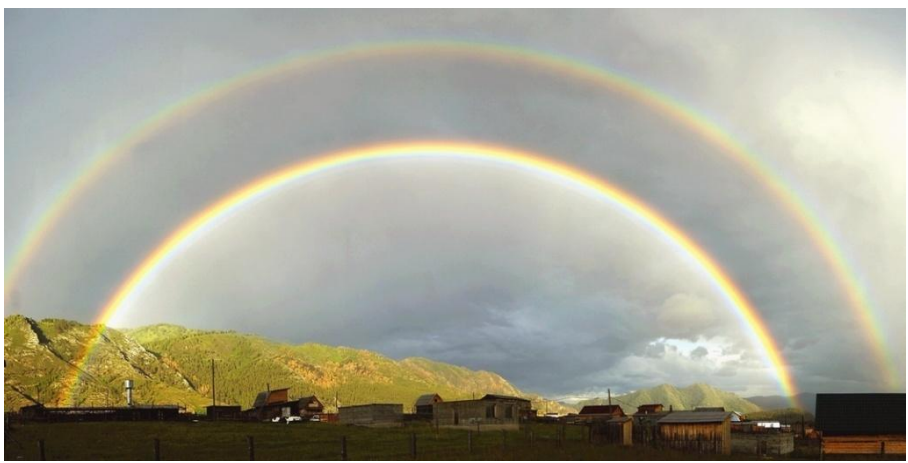

Рис. 11. Двойная радуга

И тут нужно сказать, что почти всегда вместе с *первой, основной дугой*, которую мы обсуждали и которая видна под углом 42°, появляется менее яркая, внешняя *вторая дуга*, видная под углом 51° (рис. 11). На этом рисунке также заметна темная полоса, расположенная между дугами. Она называется *темным пространством Александра*, по имени описавшего ее греческого философа-перипатетика Александра Афродисийского.

**Вторая дуга**. Вторая дуга идеально вписывается в теорию Декарта. Уже в процессе эксперимента с колбой Декарт отмечает наличие второго, менее яркого конуса с углом полураствора 51°, излучаемого каплей в направлении к солнцу. То есть рисунок 1 на самом деле нужно заменить рисунком 12 с двумя конусами.

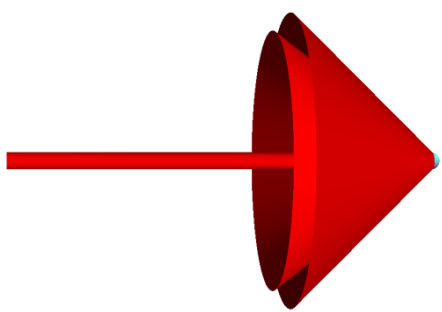

Рис. 12. В направлении к солнцу капля излучает два конуса,
один с углом полураствора 42°, другой с углом 51°

Если теперь вернуться к рисунку 2 и поменять там угол $42°$ на угол $51°$, то становится ясным, откуда берется вторая дуга.

Опять же, экспериментально Декарт установил, что за образование второй дуги отвечают лучи, дважды отразившиеся внутри капли (рис. 13).

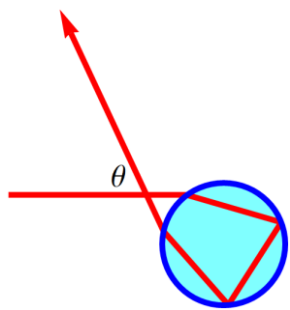

Рис. 13. Лучи дважды отражающиеся внутри капли, отвечают за образование второй дуги, $\theta$ – угол выхода

Декарт рассчитал траекторию таких лучей. Он выяснил, что выходящие из капли лучи заполняют внешнюю часть конуса с углом полураствора $51°$, при этом на поверхности конуса наблюдается повышенная концентрация световых лучей (рис. 14). Так что после некоторой корректировки рисунок 12 остается в силе. Остается в силе и объяснение второй дуги, основанное на рассмотрении рисунка 2, только, опять же, там $42°$ нужно поменять на $51°$.

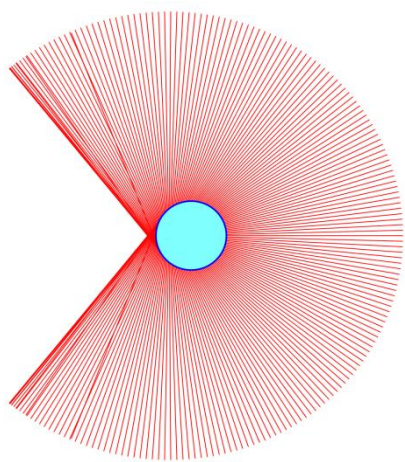

Рис. 14. Все лучи, выходящие из капли после двух внутренних отражений

**Упражнение 6**. Так же, как это было сделано для однократно отраженного внутри единичной капли светового луча в упражнениях 3 и 5, докажите, что для дважды отраженного внутри капли луча (рис. 13) выполнены следующие утверждения.
1. Зависимость угла выхода $\theta$ от угла падения $i$ и угла преломления $r$ имеет вид
$$\theta = 2i - 6r + \pi$$

2. Зависимость угла выхода θ от высоты входа $h$ имеет вид
$$\theta(h) = 2\arcsin h - 6\arcsin\left(\tfrac{3}{4}h\right) + \pi.$$
Постройте график этой функции. Докажите, что ее минимум достигается при $h = \sqrt{130}/12$, а само минимальное значение в градусах равно 50°59'.

**Пространство Александра.** Теперь наложим друг на друга рисунки 9 и 14. Рисунок 15 показывает, что на большом удалении от капли первая дуга лежит внутри второй. Их разделяет пространство, куда не попадают ни один из лучей, однократно или двукратно отраженных внутри капли. С точки зрения наблюдателя это означает, в его глаз не будут попадать лучи от капель, расположенных между первой и второй дугой. Именно за счет этого пространство между дугами выглядит более темным (еще раз посмотрите на рисунок 11).

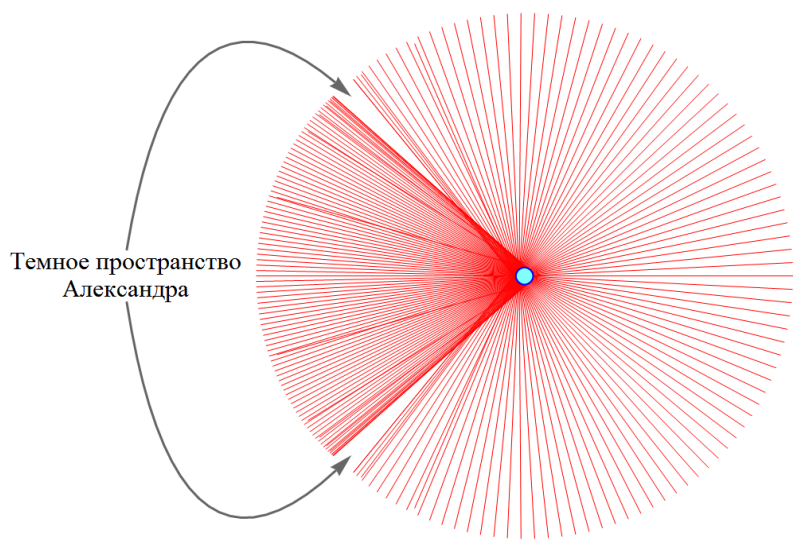

Рис. 15. Два конуса и пространство Александра

**Упражнение 7**. Оцените угловую ширину темного пространства Александра в градусах.

Сделаем небольшое дополнение. Судя по рисунку 9, вершина конуса с углом полураствора 42° лежит вблизи задней поверхности капли. А рисунок 14 показывает, что вершина конуса с углом полураствора 51° находится вблизи передней поверхности капли. Это приводит к тому, что вблизи капли больший конус расположен внутри меньшего, что также подтверждается и рисунком 15.

**Эксперимент с колбой**. Мы почти что закончили с теорией Декарта. Попытаемся еще воспроизвести его эксперимент. "Каплей" послужит круглодонная колба объемом 250 мл. На экране компьютера я нарисовал белый круг на черном фоне – изображение солнца, и запустил это изображение через проектор. Для наблюдения радуги в качестве экрана использовался лист ватмана с круговым отверстием посередине. На рисунке 16 видны колба и экран, а проектор, вместе с компьютером, находится за экраном и через отверстие освещает колбу.

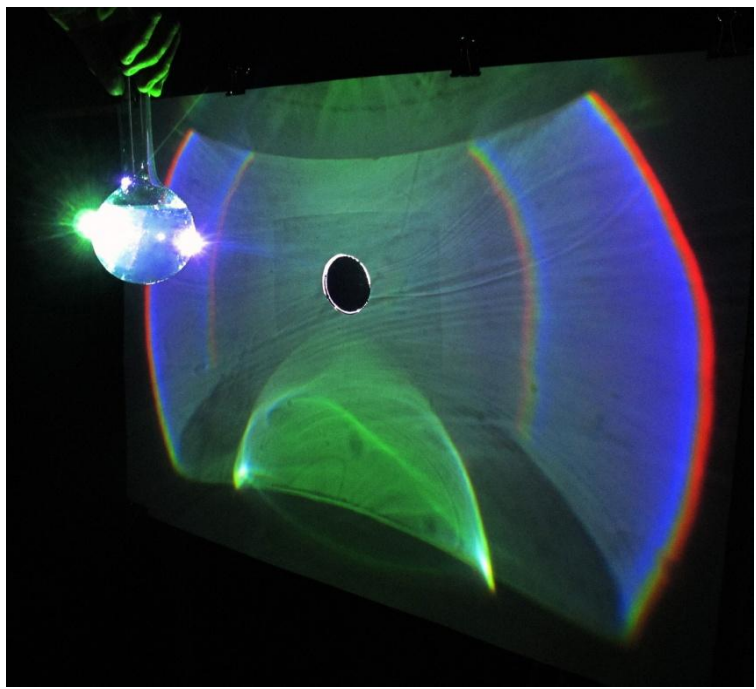

Рис. 16. Эксперимент с колбой, вторая дуга внутри первой

Колба – это конечно не сферический сосуд, у нее есть горло, так что водой была заполнена только ее сферическая часть. Поэтому круговые дуги на рис. 16 видны лишь частично, при этом нижняя и верхняя часть изображения на экране формируется горизонтальной поверхностью воды в колбе.

Где здесь первая дуга, а где вторая? Вот признаки, по которым мы можем их отличить. Вторая – менее яркая, за счет того, что формирующие ее лучи менее концентрированы (рис. 15) и к тому же испытывают внутри капли дополнительное отражение. Кроме того, в первой дуге цвета изнутри снаружи идут от фиолетового к красному, а во второй в обратном порядке. Судя по этим признакам, внутри находится вторая дуга. И это соответствует теории Декарта, помощью которой был построен рисунок 15 – там тоже вблизи капли вторая дуга расположена внутри первой. Дуги на рисунке 16 обращены друг к другу фиолетовыми сторонами и пространство между ними заполнено фиолетовым цветом. Это пространство, расположенное вблизи капли, по аналогии, можно назвать *фиолетовым пространством Александра*.

Судя по рисунку 15 стандартное взаимное расположение дуг, при котором первая лежит внутри второй, восстанавливается, когда расстояние от капли до экрана будет больше десяти радиусов капли. И конечно, при наблюдении обычной радуги расстояние от капель дождя до глаза наблюдателя многократно превышает этот предел.

## Радуга Ньютона

Хотя Декарт в своих "Рассуждениях о методе" и пытался дать объяснение цветам радуги, но эта попытка была неудовлетворительной. Решающее слово тут было сказано Ньютоном. В 1704 году вышла его знаменитая "Оптика", в начале которой он описал свои опыты с призмами, изложил свою теорию цвета и тут же использовал ее для объяснения цветов радуги.

Если на призму падает луч солнечного света, то из призмы выходит целый веер – спектр световых лучей (рис. 17). Каждый из этих лучей имеет свой цвет и преломляется призмой на свой угол. Ньютон исследует эти лучи и называет их однородными, а мы сейчас называем их монохроматическими.

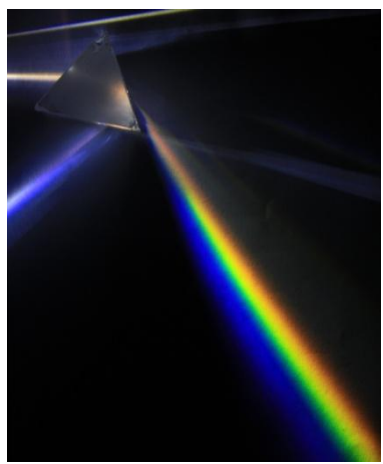

Рис. 17. Призма разлагает белый свет в целый спектр монохроматических лучей

По Ньютону луч солнечного света является смесью монохроматических лучей, каждый из которых удовлетворяет своему закону преломления. Если раньше для перехода светового луча из воздуха в воду мы записывали его в виде $\sin i = \frac{4}{3}\sin r$, то теперь нужно учитывать, что показатель преломления для воды не является постоянной величиной $n = 4/3$, а зависит от цвета монохроматического луча.

По нынешним представлениям зависимость показателя преломления воды может быть выражена графически следующим образом, как это сделано на рисунке 18. На нем горизонтальная ось имеет двойную маркировку. С одной стороны она раскрашена в цвета видимого спектра, и так мог сделать сам Ньютон. С другой стороны она размечена длинами волн, соответствующих этим цветам. Вот этого Ньютон сделать не мог, так как волновой теории света тогда еще не существовало.

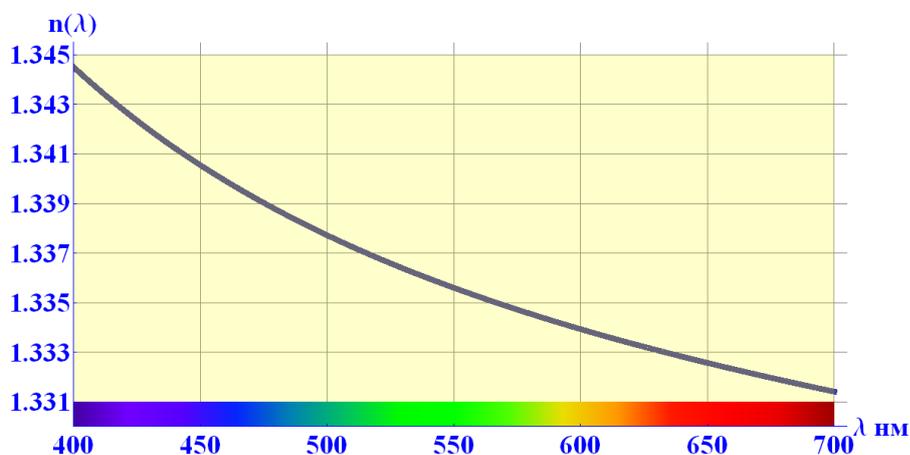

Рис. 18. Зависимость показателя преломления воды от длины волны видимого спектра и соответствующего цвета

По измерениям Ньютона показатель преломления для крайнего красного цвета света равен $n_{к} = 4/3 = 1.33333$, а для крайнего фиолетового он равен $n_{ф} = 109/81 = 1.34568$. И то и другое слегка отличается от того, что мы видим на рисунке 18.

Но теперь мы можем сказать, что все предыдущие расчеты для декартовой радуги мы проводили с лучами красного цвета с показателем преломления $n_{к} = 4/3$. Мы установили, что первая красная дуга имеет угловой радиус $42°2'$, а вторая $50°59'$ (см. упражнения 5 и 6). Это совпадает с результатами Ньютона.

А теперь все, то же самое, нужно проделать с фиолетовыми лучами. Если грубо, то на рисунках 1-9, 12-15 нужно поменять красный цвет на фиолетовый. А если быть более точными, то расчеты, проделанные в упражнениях 5 и 6 для показателя преломления $n_{к} = 4/3$, нужно повторить для показателя преломления $n_{ф} = 109/81$.

**Упражнение 7**. Вернитесь к упражнениям 5 и 6, замените там показатель преломления 4/3 на 109/81, и докажите, что угловой радиус первой фиолетовой дуги равен $40°16'$, а радиус второй фиолетовой дуги $54°10'$.

Результаты, указанные в упражнении 7, совпадают с полученными Ньютоном. Таким образом, Ньютон считает, что первая радуга заключена между фиолетовой дугой с угловым радиусом $40°16'$ и красной с радиусом $42°2'$, а вторая – между красной дугой с угловым радиусом $50°59'$ и фиолетовой с радиусом $54°10'$. Разности первых двух величин дают ширину первой радуги, разность других – ширину второй радуги (табл. 1). В первой радуге цвета изнутри наружу идут в порядке от фиолетового к красному, во второй радуге порядок цветов обратный.

|  | цвет | угловой радиус | угловая ширина |
|---|---|---|---|
| первая дуга | фиолетовый | 40°16' | 1°46' |
| | красный | 42°02' | |
| вторая дуга | красный | 50°59' | 3°11' |
| | фиолетовый | 54°10' | |

Таблица 1. Радиусы и ширина первой и второй дуг

Ньютон делает еще некоторые поправки к этим результатам, но мы на них останавливаться не будем. Итак, ширина обеих дуг вычислена, порядок цветов в каждой из них определен, и теория Декарта-Ньютона завершена.

Пожалуй, стоит еще добавить, что на самом деле Декарт проводил все свои расчеты для фиксированного показателя преломления $n = 250/187 = 1.33680$. Если посмотреть на рисунок 18, то это соответствует зеленой части спектра. Таким образом, радуга Декарта должна была выглядеть тонкой зеленой дугой окружности.

## Радуга Юнга

На самом деле и Декарт и Ньютон пропустили один существенный элемент радуги, а именно, – *дуги высших порядков*. Они довольно часто бывают видны внутри первой дуги, вплотную прилегая к ней (рис. 19). Геометрическая оптика слишком прямолинейна, чтобы дать объяснение этому явлению.

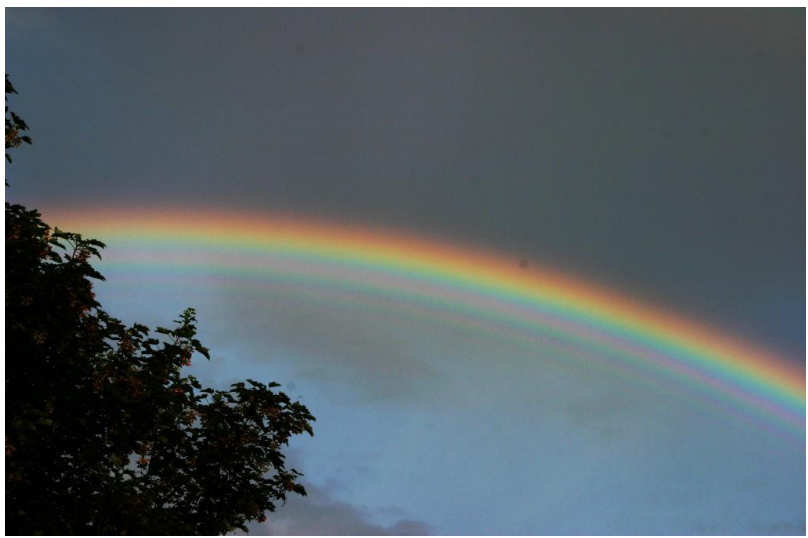

Рис. 19. Дуги высших порядков

В 1801 году Юнг провел свой знаменитый двухщелевой эксперимент, послуживший решающим аргументом в пользу волновой теории света. Одно из первых применений волновой теории – это объяснение интерференционной природы дуг высших порядков, которое содержится в его Бейкеровской лекции "Experiments and calculations relative to physical optics", прочитанной в 1803 году.

**Световые пучки и волновые фронты**. В геометрической оптике одним из основных объектов является точечный источник, от которого в разные стороны расходятся световые лучи. Что-то вроде маленького светодиода, на который мы смотрим с расстояния в несколько метров, не различая его размеров.

Важна и другая ситуация, когда источник ограниченных размеров удален от нас на гигантское расстояние. Тогда лучи, идущие от него, практически параллельны. Такой источник называется бесконечно удаленным, и пучок лучей, приходящих от него, является пучком параллельных лучей. Именно такие пучки приходят к нам от удаленных звезд, и почти что такой пучок приходит от солнца.

*Волновым фронтом* в геометрической оптике называется поверхность (в плоском случае кривая), которая перпендикулярна всем лучам из данного светового пучка. Для пучка параллельных лучей волновой фронт плоский. В случае точечного источника волновой фронт представляет собой сферу (окружность), центр которой совпадает с местом положения источника (рис. 20).

На рисунке 20 видно, что для данного светового пучка любые два фронта *эквидистантны*, то есть отрезки световых лучей, заключенные между этими фронтами, равны между собой.

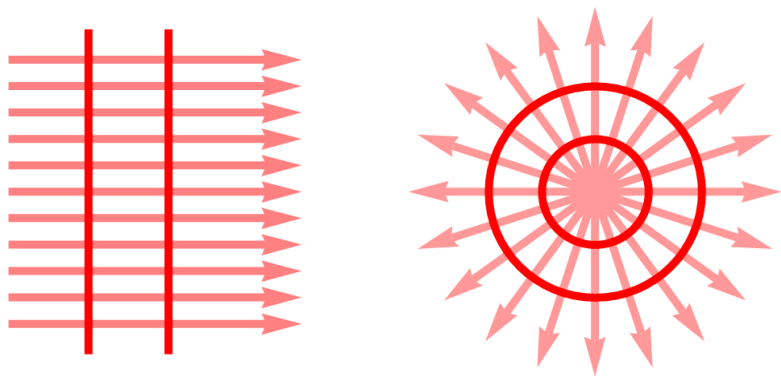

Рис. 20. Пучки и фронты перпендикулярны, фронты эквидистантны

**Монохроматическая световая волна**. Если у нас имеется монохроматический пучок световых лучей, то можно сказать, что соответствующая световая волна это совокупность волновых фронтов, удаленных друг от друга на одно и то же расстояние, равное соответствующей длине волны. Для точечного источника это будет что-то, вроде рисунка 21.

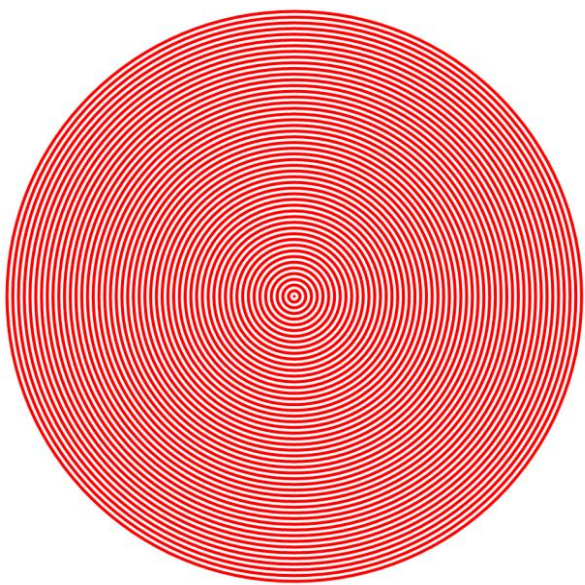

Рис. 21. Световая волна, соседние фронты эквидистантны,
расстояние между ними равно длине волны

Если считать, что световая волна распространяется в плоскости рисунка 21, то ее естественно назвать *круговой*. Если волна от точечного источника распространяется в пространстве, то это *сферическая* волна. Но на тот же рисунок можно взглянуть и по-другому. Можно считать, что начальный волновой фронт – это прямая, перпендикулярная плоскости рисунка 21 и проходящая через его центр. Тогда мы имеем дело с *цилиндрической* волной.

**Интерференция**. Рисунок 21 крайне прост, но посмотрим, что происходит при наложении двух таких рисунков со сдвигом. Иными словами, что происходит при наложении двух монохроматических волн с одной длиной волны? Картина удивительным образом преображается – мы наблюдаем интерференцию (рис. 22).

На рисунке 22 два источника смещены друг относительно друга на несколько длин волн по горизонтали. Уже при небольшом удалении точки наблюдения от обоих источников, волновые фронты приходят в нее почти параллельными. А вблизи точки, куда фронты приходят со сдвигом в целое число длин волн, они просто совпадают, и с одной стороны амплитуда волны там возрастает вдвое, с другой – вблизи этой точки на рисунке 22 мы видим много белого цвета. Окрашенные в белый цвет области – это области *конструктивной интерференции*, где интенсивность света возрастает.

Если же в некоторую точку волновые фронты приходят со сдвигом в нечетное число полуволн, то там волны гасятся, а плоскость оказывается целиком окрашенной в красный цвет (рис. 22). Мы наблюдаем области с *деструктивной интерференцией*.

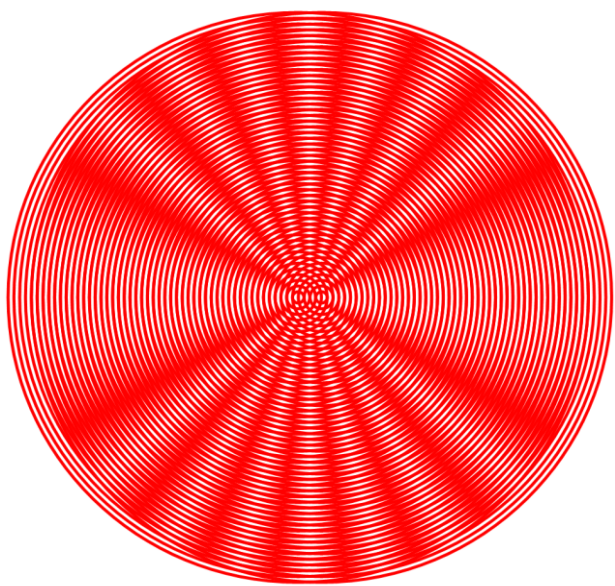

Рис. 22. Интерференция: опыт Юнга, кольца Ньютона, дуги высших порядков

Можно сказать, что на рисунке 22 области конструктивной интерференции кодируются белым цветом, а области деструктивной интерференции – красным.

**Упражнение 8**. Докажите, что на рисунке 22 области конструктивной и деструктивной интерференции расположены вдоль гипербол, фокусами которых служат два точечных источника, от которых расходятся круговые волны. Для доказательства этого факта воспользуйтесь геометрическим определением гиперболы.

В лекциях Юнга, изданных в 1807 году, есть подобный рисунок, который он использует одновременно и для описания интерференции волн на воде и для описания интерференции световых волн.

**Двухщелевой опыт Юнга и кольца Ньютона**. Это два классических интерференционных эксперимента, объяснение которым дал Юнг, и которые мы проиллюстрируем с помощью рисунка 22.

- Можно представить себе, что две узкие щели, через которые пропускал свет Юнг в своем эксперименте, – это две прямые, перпендикулярные плоскости рисунка 22, от которых распространяются две *цилиндрические* волны. В пространстве расположим плоский экран, который перпендикулярен плоскости рисунка и одновременно параллелен прямой, соединяющей источники. Цилиндрические волны интерферируют и, как показывает рисунок 22, на экране должно возникнуть чередование светлых и темных полос.
- Теперь представим себе, что мы имеем дело с двумя точечными источниками, от которых распространяются две сферические волны. Плоский экран, расположим перпендикулярно прямой, соединяющей источники. На этот раз интерферируют две *сферически* волны и, как показывает рисунок 22, на экране должно возникнуть чередование светлых и темных колец. Примерно так образуются кольца Ньютона. Только там интерферируют не расходящиеся сферические волны, а сходящиеся – каждая в своем центре.

**Радуга Юнга – дуги высших порядков**. На самом деле радуга тоже представляет собой некоторый фрагмент рисунка 22. Сейчас мы покажем это. Сначала вернемся к рисункам 7 и 8, и запустим на каплю не 101 луч, а всего 11. И сразу же нарисуем волновые фронты, *перпендикулярные* световым лучам (рис. 23). На входе в каплю, мы имеем пучок параллельных лучей и, соответственно, семейство параллельных между собой фронтов, отстоящих друг от друга на расстояние длины волны. Входя в каплю, лучи преломляются, а фронты искривляются. Причем внутри капли расстояние между фронтами уменьшается в $n$ раз, где $n = 4/3$ – показатель преломления воды, но ровно во столько же раз уменьшается и длина световой волны. Так что и в воде волновые фронты тоже эквидистантны и расстояние между ними тоже равно длине световой волны.

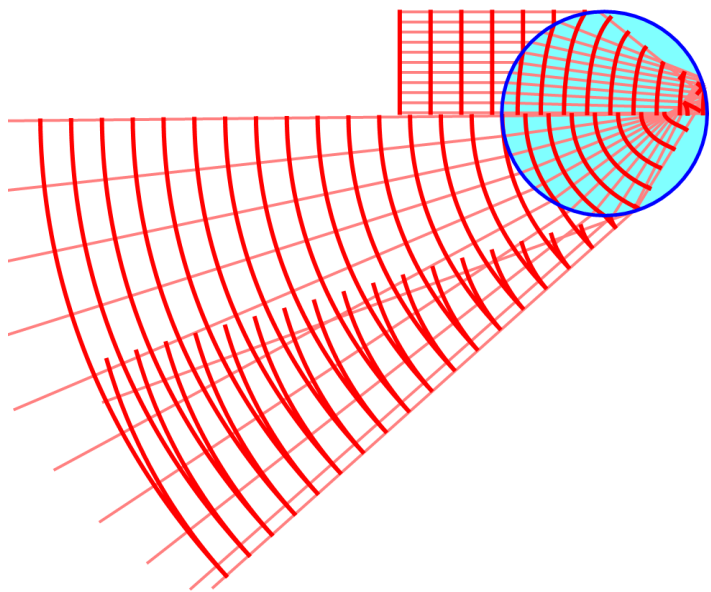

Рис. 23. Лучи и фронты перпендикулярны, фронты эквидистантны

Теперь посмотрим на волновые фронты по выходе из капли. Каждый такой фронт состоит из двух дуг (рис. 23). Большая дуга перпендикулярна тем выходящим лучам, которые имеют высоту входа в каплю от 0 до 0,86 от радиуса капли – рисунок 7. А меньшая дуга перпендикулярна выходящим лучам с высотой входа от 0,86 до 1 – рисунок 8. Система больших дуг похожа на дуги окружностей с общим центром, расположенным справа от капли. Система меньших дуг тоже похожа на дуги окружностей, но с общим центром, расположенным внизу капли. Таким образом, выходящие фронты образуют фрагмент некоторый картины, аналогичной той, что на рисунке 22, и должны интерферировать. Единственная причина, по которой мы не наблюдаем интерференцию на рисунке, – это то, что мы на рисунке взяли слишком большую длину волны.

Стоит ее уменьшить, и сразу же возникает интерференционная картина (рис. 24).

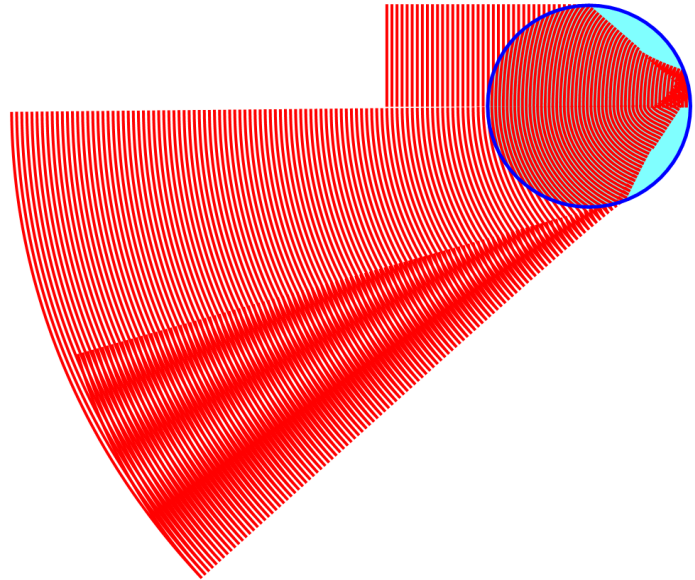

Рис. 24. Радуга Юнга

Тут светлая интерференционная полоса, прилегающая к угловым точкам волновых фронтов, идет под углом 42° и определяет направление первой дуги. Следующие светлые зоны конструктивной интерференции указывают направление на дуги высших порядков. Так что дуги высших порядков на рисунке 19 получил свое объяснение.

## Добавление

В заключение – два слова о теориях радуги, появившихся после Юнга, еще одно упражнение и еще один компьютерный рисунок.

**Теория Эйри**. В 1838 году Джорж Эйри опубликовал статью *On the intensity of light in the neighborhood of a caustic*, в которой он использовал дифракционную теорию, основанную на принципе Гюйгенса-Френеля, и создал новую теорию радуги. В ней были уточнены расположения дуг, в том числе и высших порядков. Кроме того, на границах двух основных дуг отсутствовала бесконечная интенсивность света, присущая всем предыдущим теориям, и за счет дифракции свет проникал в темное пространство Александра.

**Теория Ми**. Когда Максвелл создал теорию электромагнитного поля, выяснилось, что как раз ее и нужно использовать для точного описания радуги. А именно, нужно решить уравнения Максвелла для случая взаимодействия плоской электромагнитной волны со сферической каплей воды. Это было сделано в 1908 году почти одновременно Густавом Ми и Питером Дебаем, а намного позже выяснилось, что та же самая задача была решена Людвигом Лоренцем еще в 1890 году. Решение оказалось достаточно сложным и ему присвоили название *теория Ми*. Для реализации этого решения требовалось гигантское количество вычислений, практически не доступное до появления компьютеров. Хорошее обсуждение теории Ми есть на двух сайтах, указанных ниже в ссылках.

**Радуга Птолемея**. Теперь мы возвращаемся на много лет назад. При изложении декартовой теории радуги мы использовали закон преломления $\sin i = \frac{4}{3}\sin r$ (для лучей красного цвета, переходящих из воздуха в воду). Перепишем его в виде
$$r = \arcsin\left(\tfrac{3}{4}\sin i\right).$$

Во второй половине II века, проводя свои эксперименты, Птолемей нашел очень хорошее приближение к этому закону $r = i\left(33 - \frac{i}{10}\right)/40$, при этом он измерял углы в градусах. А мы переведем градусы в радианы, тогда закон Птолемея примет вид
$$r = i\left(\frac{33}{40} - \frac{9i}{20\pi}\right),$$
и сформулируем последнее упражнение.

**Упражнение 9**. Используя вместо точного закона преломления приближенный закон Птолемея, повторите все, что мы проделали в первой части, обсуждая радугу Декарта. Интересно, возникнет ли в этом случае первая дуга? А вторая?

**Радуга из компьютера**. Хорошим подтверждением теории Декарта-Ньютона может служить еще одна компьютерная картинка. Представьте себе, что вы смотрите на лист кальки с круглой дыркой посередине. За листом кальки находится маленькая капля, и через эту дырку она освещается пучком солнечных лучей. Что за картина возникнет на листе кальки?

А теперь представим, что капля находится не за листом кальки, а за экраном компьютера. Запустим на нее 1 000 000 монохроматических световых лучей разного цвета. Используя закон преломления и закон отражения, а также учитывая зависимость коэффициента преломления от цвета (рис. 18), мы можем рассчитать траектории этих лучей, испытавших одно или два отражения внутри капли, и найти точку пересечения каждой из рассчитанных траекторий с экраном компьютера. Отметим эту точку на экране соответствующим цветом и получим картину, представленную на рисунке 25. Она полностью соответствует теории Декарта-Ньютона.

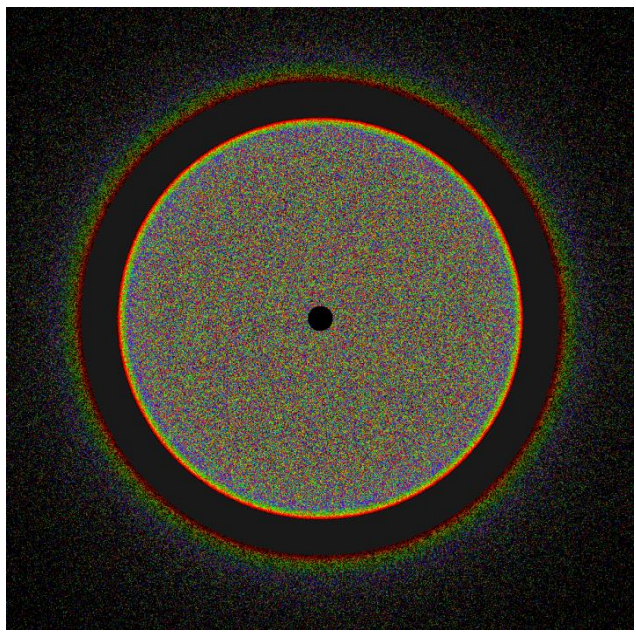

Рис. 25. Компьютерная радуга Декарта-Ньютона

**Ссылки**

Работы классиков
- Р. Декарт. *Рассуждение о методе с приложениями. Диоптрика, метеоры, геометрия*. М.: АН СССР, 1953
- И. Ньютон. *Оптика или Трактат об отражениях, преломлениях, изгибаниях и цветах света*. М: Гостехиздат, 1954
- И. Ньютон. *Лекции по оптике*. М.: АН СССР, 1946
- T. Young. Bakerian Lecture [1803]: *Experiments and calculations relative to physical optics*. Philosophical Transactions of the Royal Society 94, 1804

На русском языке также имеются
- отличная статья Х. Нуссенцвейга *Теория радуги*, Успехи физических наук, № 125, 1978
- отличная книга М. Миннарта *Свет и цвет в природе*, М.: ФМЛ, 1969

Вот еще два сайта с изложением современной теории радуги, с общедоступными программами для расчета радуги и с большим количеством картинок
- Les Cowley: *Atmospheric Optics* http://atoptics.co.uk/bows.htm
- Philip Laven: *The optics of a water drop* http://www.philiplaven.com/index1.html